\documentclass{ws-procs9x6}
\usepackage{graphicx}
\begin{document}
\title{Higher twists in spin structure functions from
a ``constituent quark'' point of view}
\author{A.~V.~Sidorov}
\address{Bogoliubov Laboratory of Theoretical Physics, 
Joint Institute of Nuclear Research, 
141980 Dubna, Russian Federation}
\author{C.~Weiss}
\address{Theory Group, Jefferson Lab, 12000 Jefferson Avenue,
Newport News, VA~23606, U.S.A.}
\maketitle
\abstracts{We discuss the implications of a ``constituent quark''
structure of the nucleon for the leading ($1/Q^2$--) power corrections 
to the spin structure functions. 
Our basic assumption is the presence of quark--gluon correlations
in the nucleon wave function, whose size, $\rho \sim 0.3\, {\rm fm}$, 
is small compared to the nucleon radius, $R$ (two--scale picture). 
We argue that in this picture the isovector twist--4 matrix element 
in the proton has a sizable negative value, 
$M_N^2 |f_2^{u - d}| \sim \rho^{-2}$,
while the twist--3 matrix elements are small, $M_N^2 d_2 \sim R^{-2}$. 
These findings are in agreement with the result of a QCD fit to the $g_1$ 
world data, including recent neutron data from HERMES and 
Jefferson Lab Hall A, which gives 
$M_N^2 f_2^{u - d} = -0.28 \pm 0.08 \; {\rm GeV}^2$.}
The transition from the scaling regime at large $Q^2$ to the 
quasi--real regime at small $Q^2$ in the structure
functions of inelastic $eN$ scattering represents 
a major challenge for theory and experiment. 
Coming from high $Q^2$, the onset of the transition manifests itself 
in power ($1/Q^2$--) corrections to the $Q^2$ dependence of the structure
functions. In QCD, these corrections are related to the interactions of the 
``active'' quark/antiquark with the non-perturbative gluon field 
in the nucleon, described by nucleon matrix elements of certain 
quark--gluon operators of twist 3 and 4. What are the scale parameters 
governing the size of these matrix elements?

There is ample evidence for a constituent quark structure of the nucleon 
--- the presence in the nucleon of small--size extended objects --- 
from hadron spectroscopy and low--energy electromagnetic interactions. 
The notion of a massive constituent quark of finite size is also 
intimately related to the spontaneous breaking of chiral symmetry. 
For instance, the microscopic picture of chiral symmetry breaking 
based on QCD instantons gives rise to constituent quarks of a
``size'' of $\rho \sim 0.3 \, {\rm fm}$, which is determined by
the average size of the instantons in the vacuum.\cite{Diakonov:2002fq}
It is natural to ask if this scale could explain the quark--gluon
correlations giving rise to power corrections to polarized 
deep--inelastic scattering.\footnote{Further evidence for 
constituent quarks of a size $\rho \sim 0.3 \, {\rm fm}$ 
comes from the correlations in the transverse spatial distribution 
of partons in the nucleon, observed in the production of multiple 
hard dijets in high--energy $pp$ collisions.\cite{fest}}

In this note\cite{paper} we explore the implications of a constituent 
quark structure of the nucleon for the leading ($1/Q^2$--)
power corrections to the nucleon spin structure functions,
from a general, model--independent perspective. We think of 
constituent quarks and antiquarks in the field--theoretical sense, 
as finite--size correlations in the quark--gluon wave function 
of the nucleon in QCD, not as elementary objects in the sense 
of a potential model. 
We shall try to relate the size of these correlations to the
nucleon matrix elements of twist--3 and 4 quark--gluon operators
which govern the $1/Q^2$--corrections to the lowest moments
of the spin structure functions in QCD.

\begin{figure}[t]
\begin{center}
\includegraphics[width=3cm,height=3cm]{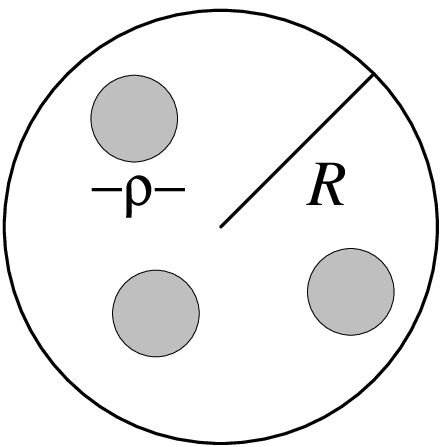}
\end{center}
\caption[*]{}
\label{fig_twoscale}
\end{figure}
Our basic assumption is that the size of the ``constituent quarks'',
$\rho$, be much smaller than the radius of the nucleon, $R$
(see Fig.~\ref{fig_twoscale}),
\begin{equation}
\rho \;\; \ll \;\; R .
\label{hierarchy}
\end{equation}
Various phenomenological considerations point to a constituent quark 
size of $\sim 0.3\; \mbox{fm}$, which should be compared, say, to the nucleon
isoscalar charge radius, $\langle r^2 \rangle^{1/2} = 0.8 \; {\rm fm}$. 
The precise values of these parameters are not the issue here;
what is important is that we have a two--scale picture.
We stress that the hierarchy (\ref{hierarchy}) is really a logical 
necessity --- if the size of the constituent quark were comparable 
to that of the nucleon, we would not be ``seeing it'' 
as an independent dynamical entity. 

In QCD, the leading ($1/Q^2$) power corrections to the lowest
moments of the spin structure functions $g_1$ and $g_2$ 
are governed by the matrix elements of twist--3 and 4 operators
which measure non-perturbative correlations of the quark and gluon fields 
in the nucleon (for details, see 
Refs.\cite{Shuryak:1981pi,Ji:1994sv,Balla:1997hf}):
\begin{equation}
\begin{array}{lccl}
d_2 : \;\;\;\; & \bar\psi 
\gamma_{\left\{\alpha\right.} \widetilde F_{\left.\beta\right\} \gamma} 
\psi & \phantom{xxxx} & \mbox{Twist--3}
\\[.5cm]
f_2 : \;\;\;\; & \bar\psi \gamma_\alpha \widetilde F_{\beta\alpha} \psi &
& \mbox{Twist--4}
\end{array}
\label{ope}
\end{equation}
where $\widetilde F_{\alpha\beta} = (1/2) 
\epsilon_{\alpha\beta\gamma\delta} F_{\gamma\delta}$ is the dual
gluon field strength. With the help of the well--known relation
\begin{equation}
F_{\alpha\beta} \;\; = \;\; 
i \left[ \nabla_\alpha , \nabla_\beta \right],
\hspace{3em} \nabla_\alpha \equiv \partial_\alpha - i A_\alpha 
\;\;\; \mbox{covariant derivative},
\label{identity}
\end{equation}
and making use of gamma matrix identities and the 
equations of motion of the quark fields, the twist--4 operator
can equivalently be expressed as
\begin{equation}
\bar\psi \gamma_\beta \gamma_5 (-\nabla^2 ) \psi .
\label{axial_virtuality}
\end{equation}
In this form, it can be compared with the axial current operator, 
which measures the quark contribution to the nucleon spin,
\begin{equation}
g_A : \;\;\;\; \bar\psi \gamma_\beta \gamma_5 \psi .
\label{axial}
\end{equation}
We see that the operator (\ref{axial_virtuality}) measures
the correlation of the spin of the quarks with their virtuality
(four--momentum squared, $k^2$) in the nucleon.

\begin{figure}[t]
\begin{center}
\includegraphics[width=8.0cm,height=4cm]{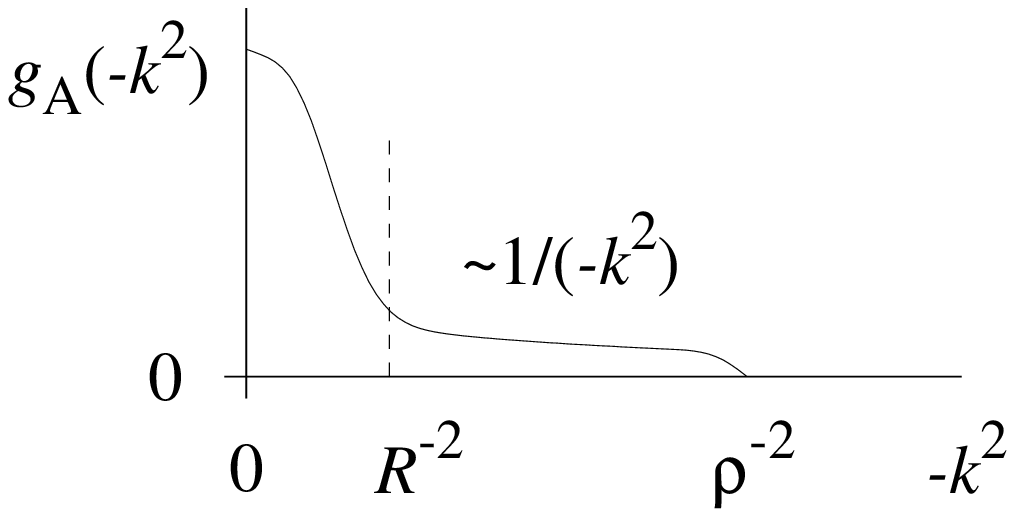}
\end{center}
\caption[*]{}
\label{fig_gadist}
\end{figure}
The constituent quark picture implies that,
generally speaking, the distribution of virtualities of quarks in
the nucleon has two components. The bulk
of the distribution is governed by the size of the nucleon,
$-k^2 \sim R^{-2}$. In addition, there is a ``tail'' extending up to 
values of the order of the inverse size of the constituent 
quark, $-k^2 \sim \rho^{-2}$. This two--component structure, which 
follows from the Fourier image of the two--scale picture of 
Fig.~\ref{fig_twoscale}, is the key to our estimates of 
higher--twist matrix elements.

Specifically, let us consider the distribution of quark virtualities
in the proton's isovector (flavor--nonsinglet) axial charge, $g_A$,
shown schematically in Fig.~\ref{fig_gadist}. In the isovector case one 
can argue that the large--virtuality tail of the distribution
is of positive sign and decays as $1/(-k^2)$, 
until it is ``cut off'' by the constituent quark size, $\rho$. 
This follows from the requirement that in the limit of small size of the
constituent quark, $\rho \ll R$, the axial charge should exhibit the
logarithmic divergence it has in QCD (with $\rho^{-1}$ acting as the 
ultraviolet cutoff). We note that exactly this behavior is found 
also in a field--theoretical chiral model in which the constituent 
quarks/antiquarks couple to a pion field.\footnote{In the isoscalar
case the behavior is different, due to the presence of the 
$U(1)$ anomaly. The following arguments do not apply in this case.} 
Thus, the isovector axial charge, 
which is the integral of the distribution shown in Fig.~\ref{fig_gadist}, 
behaves parametrically as
\begin{equation}
g_A \;\; = \;\; \int_0^\infty d(-k^2) \; g_A (-k^2) 
\;\; \sim \log \frac{\rho}{R} .
\label{g_A_integral}
\end{equation}
This integral is dominated by virtualities
$-k^2 \sim R^{-2} \ll \rho^{-2}$. Consider now the corresponding 
integral for the isovector (flavor--nonsinglet) twist--4 matrix 
element, $f_2^{u - d}$. Since the operator 
(\ref{axial_virtuality}) involves an additional contracted
derivative, this quantity is determined by the integral
with an additional factor $k^2$, which is parametrically of the
order
\begin{equation}
M_N^2 f_2^{u - d} 
\;\; = \;\; \int_0^\infty d(-k^2) \; g_A (-k^2) \; k^2
\;\; \sim \rho^{-2} .
\label{f_2_integral}
\end{equation}
Thus, the isovector twist--4 matrix element in our constituent
quark picture is governed by the size of the constituent quark. 
Furthermore, since $k^2 < 0$ in the integral (\ref{f_2_integral}),
we can say that
\begin{equation}
f_2^{u - d} \;\; < \;\; 0 .
\label{f_2_sign}
\end{equation}
To summarize, the constituent quark picture suggests a sizable
negative value for $M_N^2 f_2^{u - d}$ in the proton, 
of the order $\rho^{-2}$.
It is interesting that the estimates of $f_2^{u - d}$ obtained in 
various QCD--based approaches are in qualitative agreement with
this prediction, see Table~\ref{table}. The QCD sum rule estimates
of Refs.\cite{Balitsky:1990jb,Stein:1995si} 
as well as the instanton vacuum estimate of
Ref.\cite{Balla:1997hf} both give negative values of the order of 
$\sim 0.1 - 0.3 \; {\rm GeV}^2$. These results support
the constituent quark interpretation of 
higher--twist effects.
\begin{table}[ht]
\begin{tabular}{l|c} 
                     & $M_N^2 f_2^{u - d} \, [{\rm GeV}^2]$ \\ \hline
QCD sum rules (Balitskii et al.) \cite{Balitsky:1990jb} & -0.18  \\ \hline
QCD sum rules (Stein et al.) \cite{Stein:1995si}        & -0.06  \\ \hline
Instantons \cite{Balla:1997hf}                          & -0.22  \\ \hline
Bag model \cite{Ji:1994sv}                              & +0.1   \\ \hline
\end{tabular}
\caption[]{}
\label{table}
\end{table}

The only exception in Table~\ref{table} is the bag model, which 
gives a positive result for $f_2^{u - d}$. This model, however,
does not respect the QCD equations of motion [{\it i.e.}, the two
forms of the QCD operator, (\ref{ope}) and (\ref{axial_virtuality}),
would give inequivalent results], and therefore cannot claim to give a 
realistic description of quark--gluon correlations in the nucleon.

When applying the same reasoning as above to the twist--3 matrix element,
$d_2$, we find that after the substitution (\ref{identity}) 
the quark--gluon operator does not produce a contracted covariant 
derivative. In this operator, all derivatives are ``kinematical'', {\it i.e.},
they are needed to support the spin of the matrix element.
This operator does not probe the virtuality of the quarks.
Its matrix element is not determined by the size of the constituent
quark, but by the size of the nucleon, 
\begin{equation}
M_N^2 d_2 \;\; \sim \;\; R^{-2} .
\end{equation}
Thus, we find that our two--scale picture implies 
(see also Ref.~\cite{Wandzura:1977qf})
\begin{equation}
\begin{array}{ccc}
|d_2| & \;\;\; \ll \;\;\; & |f_2| .\\
\mbox{Twist--3} & & \mbox{Twist--4}
\end{array}
\label{d_2_vs_f_2}
\end{equation}

It is interesting to see to which extent the qualitative predictions
of the constituent quark picture are supported by the experimental data.
The twist--3 matrix element, $d_2$, can be extracted in a model--independent
way from the spin structure function $g_2$ (with the
Wandzura--Wilczek contribution subtracted). The SLAC E155X 
experiment\cite{Anthony:2002hy} and the recent Jefferson Lab Hall A 
analysis\cite{Zheng:2004ce} report values of $d_2^{p, n}$
of few times $10^{-3}$, which are more than an order of magnitude 
smaller than the theoretical estimates for $f_2^{u - d}$ given 
in Table~\ref{table} and our estimate for $f_2^{u - d}$ from $g_1$ 
data (see below), in agreement with the parametric ordering implied 
by the constituent quark picture (\ref{d_2_vs_f_2}).

The twist--4 matrix element, $f_2$, can only be extracted from 
the power corrections to the $Q^2$--dependence of the structure 
function $g_1$. The QCD expression for the $1/Q^2$
corrections to the first moment contains the sum of the 
matrix elements $d_2$ and $f_2$; however, $d_2$ is known 
from independent measurements. In fact, the parametric ordering
suggested by the constituent quark picture implies that
one should ascribe the $1/Q^2$ corrections of the first moment
of $g_1$ entirely to the twist--4 matrix element, $f_2$.

\begin{figure}[t]
\includegraphics[width=9.4cm,height=7.5cm]{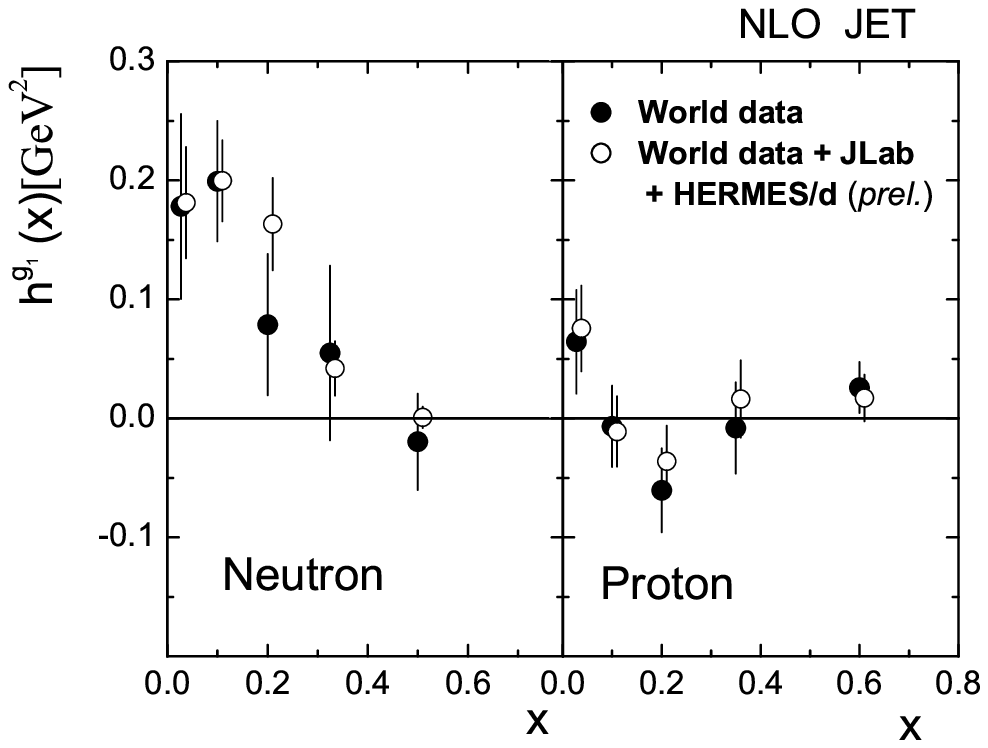}
\caption[*]{}
\label{fig_fit}
\end{figure}
The dynamical higher--twist contribution has been extracted
from NLO QCD fits to the world data on the structure functions 
$g_1^p (x, Q^2)$ and $g_1^n(x, Q^2)$ (with $Q^2 \geq 1\; 
{\rm GeV}^2$).\cite{Leader:2002ni,Leader:2003vv} They are based on the ansatz
\begin{equation}
g_1 (x, Q^2) \;\; = \;\; g_1 (x, Q^2)_{\mbox{{\scriptsize LT + TMC}}}
\; + \; \frac{h^{g_1} (x)}{Q^2} ,
\label{g_1_QCD}
\end{equation}
where the leading--twist contribution (including target 
mass corrections) is calculated using
the Leader--Stamenov--Sidorov parametrization 
of polarized parton densities,
and $h^{g_1} (x)$ parametrizes the dynamical higher--twist 
corrections. The results of Ref.~\cite{Leader:2002ni} for $h^{g_1} (x)$ 
for proton and neutron are shown in Fig.~\ref{fig_fit} 
(filled circles). The second fit\cite{Leader:2003vv} (open circles) 
includes also the new $g_1^n$ data from Jefferson 
Lab Hall A\cite{Zheng:2003un}, as well as the preliminary 
deuteron data from HERMES. One sees that the results for $h^{g_1} (x)$ 
obtained with the two data sets are nicely consistent. The new data 
allow to significantly reduce the statistical uncertainty in the
higher--twist contribution. Integrating the higher--twist
contribution over $x$ we get
\begin{equation}
\int_0^1 dx \; h^{g_1} (x) \;\; = \;\;
\left\{\begin{array}{l}
0.007 \pm 0.010 \; {\rm GeV}^2 \;\; \mbox{(proton)}
\\[0cm]
0.049 \pm 0.007 \; {\rm GeV}^2  \;\; \mbox{(neutron)}
\end{array}
\right.
\end{equation}
Comparing with the QCD expression, 
neglecting the twist--3 contribution, we obtain
\begin{equation}
M_N^2 f_2^{u - d} \;\; = \;\; -0.28 \pm 0.08 \; {\rm GeV}^2 .
\end{equation}
This estimate agrees both in sign and in order--of--magnitude
with the predictions of the constituent quark picture,
(\ref{f_2_integral}) and (\ref{f_2_sign}). 

Our result agrees well with that obtained by 
Deur {\it et al.}\cite{Deur:2004ti} in a recent analysis of 
power corrections to the Bjorken sum rule 
(their $f_2^{p - n} \equiv \frac{1}{3} f_2^{u - d}$ in our conventions).
It disagrees in sign with the result of Kao {\it et al.}\cite{Kao:2003jd}, 
who use a resonance--based parametrization of the structure function 
at low $Q^2$. The origin of this discrepancy remains to be understood.

The constituent quark picture allows to draw some interesting
conclusions about the ``global'' properties of the transition from 
high to low $Q^2$ in the nucleon spin structure functions 
({\it i.e.}, going beyond the leading $1/Q^2$ corrections).
Since the characteristic mass scale for the power corrections 
is set by the size of the constituent quark, 
one should expect the twist expansion
to break down at momenta of the order $Q^2 \sim \rho^{-2}$.
For the extraction of the leading ($1/Q^2$--) corrections from
QCD fits to the data this implies that one should restrict oneself
to the range $Q^2 \gg \rho^{-2}$, where the leading term in the 
series dominates. Physically speaking, in this region the scattering
process takes place ``inside'' the constituent quark.

When $Q^2$ is decreased to values of the order $R^{-2}$, the scattering
process probes the motion of the constituent quarks in the nucleon.
This is the region dominated by nucleon resonances. In the constituent
quark picture, these are changes of the state of motion of the 
constituent quarks at the scale $R$, which do not affect the
internal structure of the constituent quark at the scale $\rho$. 
Thus, our two--scale picture implies a clear distinction between
resonance and higher--twist contributions. It is close in spirit 
to the parametrization of the $Q^2$ dependence of the first moment
of $g_1$ by Ioffe and Burkert\cite{Burkert:1992tg}, in which the contribution
from the Delta resonance is separated from the continuum, and
the leading power corrections are associated with the 
continuum contribution.\footnote{It is interesting to note that the
mass scale governing the power corrections in the Ioffe--Burkert 
parametrization\cite{Burkert:1992tg}, $M_\rho^2$, 
is numerically close to value associated 
with the constituent quark size, 
$\rho^{-2} \sim (0.3 \, {\rm fm})^{-2} = (600 \, {\rm MeV})^2$.}

To summarize, we have shown that the assumption of a two--scale 
``constituent quark structure'' of the nucleon implies certain 
qualitative statements about the twist--3 and 4 matrix elements, 
which are in agreement with present polarized DIS data. 
For more quantitative estimates, one eventually has to turn 
to dynamical models. A consistent realization of the scenario developed 
here is provided by the instanton model of the QCD vacuum, in which the
hierarchy (\ref{hierarchy}) follows from the diluteness of the instanton 
medium.\cite{Balla:1997hf} In particular, this model
incorporates the chiral dynamics at the scale $R$, {\it i.e.}, the binding
of the constituent quarks and antiquarks in the 
nucleon\cite{Diakonov:2002fq}, and thus can serve as the basis of an 
``interpolating'' description connecting the scaling region at 
large $Q^2$ with the photoproduction point.

Finally, the constituent quark picture suggested here
can be applied also to the power corrections to the 
unpolarized structure functions; the relation of our approach
to that of Petronzio {\it et al.}\cite{Petronzio:2003bw} 
will be discussed elsewhere\cite{paper}.

This work was supported by U.S.\ Department of Energy contract \\
DE-AC05-84ER40150, under which the Southeastern Universities Research 
Association (SURA) operates the Thomas Jefferson National 
Accelerator Facility. A.~S.\ acknowledges financial support by 
RFBR (No 02-01-00601, 03-02-16816), INTAS 2000 (No 587), 
and the Heisenberg--Landau program.
The work of C.~W.\ was partly supported by DFG (Heisenberg Fellowship).
\end{document}